# A Fully Decentralized Control of Grid-Connected Cascaded Inverters

Yao Sun, Xiaochao Hou, Hua Han, Zhangjie Liu, Wenbin Yuan and Mei Su

*Abstract*—This letter proposes a decentralized control scheme for grid-connected cascaded modular inverters without any communication, and each module makes decisions based on its own local information. In contrast, the conventional methods are usually centralized control and depend on a real-time communication. Thus, the proposed scheme has advantages of improved reliability and decreased costs. The overall system stability is analyzed, and the stability condition is derived as well. The feasibility of the proposed method is verified by simulation.

## I. Introduction

Multiple inverters are widely connected in cascaded-type manners to form a high voltage level power network [1]-[5]. These cascaded-type inverters are early applied to multilevel topologies [1], and extended into distributed generations, especially for PV grid-connected system [2], micro-grids [3] and battery balance managements [4]. Most power balance methods of cascaded-type system are centralized with high bandwidth communications in grid-connected mode, which severely limits its application scope [1]-[4]. For cascaded micro-converters, [5] firstly proposes a novel power balance control with low bandwidth communications. The central controller of string converters interconnects the local controllers of micro-converters by transferring slow dynamic DC components. Though excellent power balance and voltage quality are achieved, they do not go beyond a communication-based category.

Cascaded inverters have distinct merits in grid-connected mode, but the power balance among all modules has not been researched in decentralized manners. In this letter, we introduce a $P$-$\omega$ droop control for frequency synchronization and power balance autonomously. The stability of the proposed scheme is proved in theory and the effectiveness is verified by simulation results.

## II. Analysis of Decentralized Power Balance Control

### A. Equivalent Models of Grid-Connected Cascaded Inverters

Fig. 1 illustrates the schematic diagram of grid-connected cascaded-type inverters. This configuration is beneficial to integrate low-voltage DC distributed generations (DGs) into medium voltage system. It is very common in PV grid-connected applications.

From Fig.1, the output real power $P_i$ and reactive power $Q_i$ of $i$-th module are derived as follows

$$P_i + jQ_i = V_i e^{j\delta_i} \cdot \left( (V_p e^{j\delta_p} - V_g e^{j\delta_g}) / (|Z_{line}| e^{j\theta_{line}}) \right)^* \quad (1)$$

where $V_i$ and $\delta_i$ represent the output voltage amplitude and phase angle of $i$-th module. $V_g$ and $\delta_g$ are the voltage amplitude and phase angle of utility grid. $|Z_{line}|$ and $\theta_{line}$ are the grid impedance amplitude and angle. Usually, the grid impedance is mainly inductive ($\theta_{line} \approx \pi/2$). The voltage $V_p e^{j\delta_p}$ at point of common coupling (PCC) is the voltage sum of all modules.

$$V_p e^{j\delta_p} = \sum_{j=1}^{N} V_j e^{j\delta_j} \quad (2)$$

where $N$ represents the total number of cascaded modules.

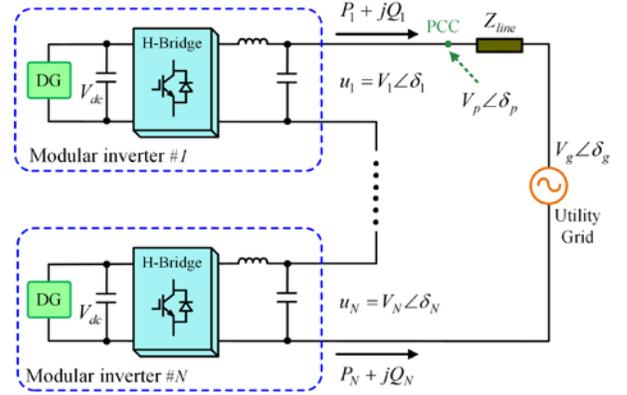

Fig. 1. Schematic diagram of grid-connected cascaded-type inverters.

From (1)-(2), the power transmission characteristic is given

$$P_i = \frac{V_i}{|Z_{line}|} \left( V_g \sin(\delta_i - \delta_g) - \sum_{j=1}^{N} V_j \sin(\delta_i - \delta_j) \right) \quad (3)$$

$$Q_i = \frac{V_i}{|Z_{line}|} \left( \sum_{j=1}^{N} V_j \cos(\delta_i - \delta_j) - V_g \cos(\delta_i - \delta_g) \right) \quad (4)$$

### B. Proposed P-$\omega$ Droop Control

To synchronize each module with the grid and realize power balance without communication, a decentralized control scheme is designed as

$$\omega_i = \omega^* - k \cdot (P_i - P^*) \quad (5)$$

$$V_i = V^* = V_g / M \quad (6)$$

where $\omega_i$ and $V_i$ are the angular frequency and voltage amplitude references of $i$-th module, respectively. $\omega^*$ represents the nominal value of the grid angular frequency. $P^*$ represents the nominal rated power of each module. $k$ is a positive coefficient of $P$-$\omega$ droop control. $M$ is a critical parameter related with stability, which is designed later.

### C. Steady-State Analysis

In steady state, because the voltage amplitude reference $V^*$ is same for all modules and each module shares the same grid current, the apparent power of each module is equal. Moreover, (7) is obtained due to the identical grid frequency from (5)

$$P_1 = P_2 = \cdots = P_N = P^* \quad (7)$$

Namely, the phase angles of all modules are equal in steady state ($\delta_i = \delta_j$; $i, j \in \{1, 2\ldots N\}$). Thus, the voltage amplitude and phase angle of PCC is derived from (2)

$$V_p = NV^* = (N/M) \cdot V_g; \quad \delta_p = \delta_1 = \delta_2 \cdots = \delta_N \quad (8)$$

Then, the steady state power of each module is expressed from (3)-(6)

$$P^* = S_C \sin \bar{\delta}; \quad \bar{Q}_i = S_C (N/M - \cos \bar{\delta}) \quad (9)$$

where $S_C = V_g^2 / (M \cdot |Z_{line}|)$ represents the power transfer capacity of a single module. $\bar{\delta} = \delta_p - \delta_g$ is referred to as the steady power angle. From (9), there are two steady points and they are $\bar{\delta} = \arcsin(\frac{P^*}{S_C})$ or $\bar{\delta} = \pi - \arcsin(\frac{P^*}{S_C})$.

### D. Stability Analysis

In this section, the small signal analysis is carried out to test the system stability. Since $\dot{\delta}_i = \omega_i$, combining (3)-(6) and linearizing them around the steady state points yield

$$\dot{\tilde{\delta}}_i = -k' \left( M(\cos \bar{\delta})(\tilde{\delta}_i - \tilde{\delta}_g) - \sum_{j=1, j \neq i}^{N} (\tilde{\delta}_i - \tilde{\delta}_j) \right) \quad (10)$$

where $k' = kV^{*2}/|Z_{line}|$, and $\tilde{\delta}_i, \tilde{\delta}_g, \tilde{\delta}_j$ mean small perturbations around the equilibrium point.

Rewrite (10) in matrix form as

$$\dot{\tilde{\delta}} = -k' \cdot L \cdot \tilde{\delta} \quad (11)$$

where $\tilde{\delta} = diag[\tilde{\delta}_1 \; \tilde{\delta}_2 \; \cdots \; \tilde{\delta}_N]$;

$$L = \begin{bmatrix} M\cos\bar{\delta}-N+1 & 1 & \cdots & 1 \\ 1 & M\cos\bar{\delta}-N+1 & \cdots & 1 \\ \vdots & \vdots & \ddots & \vdots \\ 1 & 1 & \cdots & M\cos\bar{\delta}-N+1 \end{bmatrix} = (M\cos\bar{\delta}-N)I_{N\times N} + 1_N 1_N^T$$

The eigenvalues of the system matrix $A = -k'L$ are given by

$$\lambda_1(A) = -k'M\cos\bar{\delta}; \quad \lambda_2(A) = \cdots = \lambda_N(A) = -k'(M\cos\bar{\delta}-N). \quad (12)$$

Thus, the necessary and sufficient condition of system stability is obtained as follows

$$\Delta = M\cos\bar{\delta} - N > 0 \quad (13)$$

From (13), since both $M$ and $N$ are greater than zero, the power angle $\bar{\delta}$ should lie in $(-\pi/2, \pi/2)$ under the stability constraint. According to the above steady-state analysis, only the first equilibrium point $\bar{\delta} = \arcsin(P^*/S_C)$ is feasible in (13). From (9), to obtain a high power factor, $\Delta$ should be as small as possible. However, a too small $\Delta$ is detrimental to stability from (12). Thus, a proper $M$ should be designed by making a tradeoff between reactive power requirement and stability.

### III. SIMULATION RESULTS

The simulation tests are carried out to validate the proposed ideas. The simulation parameters of the system comprised of six cascaded modules are listed in Table I. According to the steady-state analysis (9) and stability condition (13), three cases are considered. Compared with the unstable case-1 in Fig. 2(a.1)-(b.1), the stable results of case-2 in Fig. 2(a.2)-(c.2) reveal that the system can achieve real-power/reactive-power balance and frequency synchronization in steady state. Moreover, compared with the case-2, Fig. 2(a.3)-(c.3) of case-3 show that a large $M$ can result in a shorter stable settling time, but the power factor is declined from 0.983 to 0.891 in Fig. 2(d).

TABLE I
SIMULATION PARAMETERS

| Symbol | Value | Symbol | Value |
|---|---|---|---|
| $V_g$ | 311 V | | 5.8 (Case-1) |
| $\omega^*$ | 2π*50 rad/s | $M$ | 6.2 (Case-2) |
| $P^*$ | 4 kW | | 7.0 (Case-3) |
| $k$ | 1.2e-3 | | 53.6 V (Case-1) |
| $Z_{line}$ | 0.1+j0.5 Ω | $V^*$ | 50.1 V (Case-2) |
| $N$ | 6 | | 44.4 V (Case-3) |

### IV. CONCLUSION

As an important supplementary, this letter presents a fully decentralized control for cascaded inverter system in grid-connected mode. It realizes accurate power balance and frequency synchronization autonomously without relying on any communication. Based on this fundamental study, some novel hierarchical control schemes would be constructed for future PV and storage cascaded systems, where decentralized control takes a role of primary control. For readers, this letter explores the possibilities of inspiring new methods in a hybrid power network with cascaded-type and parallel-type inverters.

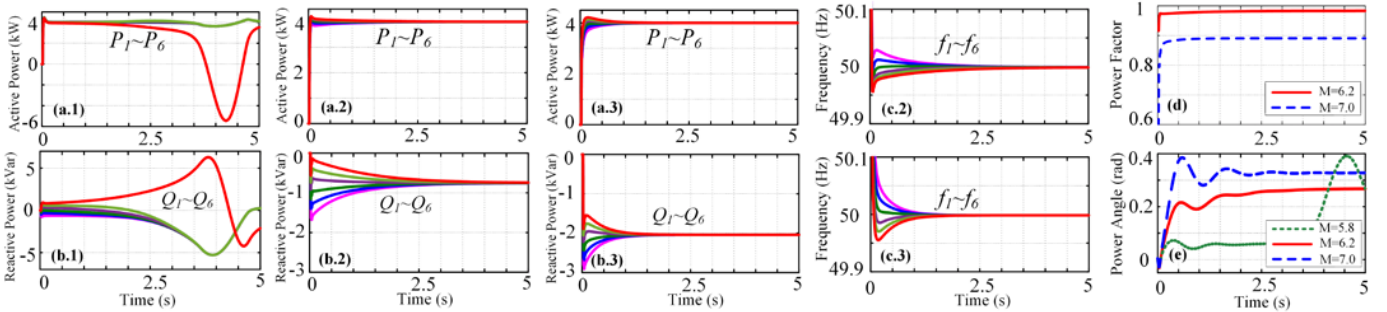

Fig. 2. Simulation results of (a) active power, (b) reactive power, (c) frequency, (d) power factor, and (d) power angle under three cases ($M=5.8$, $M=6.2$, $M=7.0$).